\documentclass[fleqn,usenatbib]{mnras}

\usepackage{newtxtext,newtxmath}

\usepackage[T1]{fontenc}

\DeclareRobustCommand{\VAN}[3]{#2}
\let\VANthebibliography\thebibliography
\def\thebibliography{\DeclareRobustCommand{\VAN}[3]{##3}\VANthebibliography}

\usepackage{graphicx}	
\usepackage{amsmath}	

\usepackage[usenames,dvipsnames]{xcolor}

\def \d {\mathrm{d}}

\renewcommand{\vec}[1]{\boldsymbol{\mathbf{#1}}} 
\renewcommand{\vec}[1]{\boldsymbol{\mathbf{#1}}}

\title[Streaming instability and pressure anisotropies]{Interplay between the non-resonant streaming instability and self-generated pressure anisotropies}

\author[A. Marret et al.]{
A. Marret,$^{1,2,3}$\thanks{E-mail: alexis.marret@slac.stanford.edu}
A. Ciardi,$^{2}$
R. Smets$^{3}$
\\
$^{1}$High Energy Density Science Division, SLAC National Accelerator Laboratory, Menlo Park, California 94025, USA\\
$^{2}$Sorbonne Universit\'e, Observatoire de Paris, Universit\'e PSL, CNRS, LERMA, F-75005, Paris, France\\
$^{3}$Sorbonne Universit\'e, Ecole Polytechnique, CNRS, Observatoire de Paris, LPP, F-75005, Paris, France
}

\date{Accepted XXX. Received YYY; in original form ZZZ}

\pubyear{2023}

\begin{document}
\label{firstpage}
\pagerange{\pageref{firstpage}--\pageref{lastpage}}
\maketitle

\begin{abstract}
The non-thermal particles escaping from collisionless shocks into the surrounding medium can trigger a non-resonant streaming instability that converts parts of their drift kinetic energy into large amplitude magnetic field perturbations, and promote the confinement and acceleration of high energy cosmic rays. We present simulations of the instability using an hybrid-Particle-in-Cell approach including Monte Carlo collisions, and demonstrate that the development of the non-resonant mode is associated with important ion pressure anisotropies in the background plasma. Depending on the initial conditions, the anisotropies may act on the instability by lowering its growth and trigger secondary micro-instabilities. 
Introducing collisions with neutrals yield a strong reduction of the magnetic field amplification as predicted by linear fluid theory. In contrast, Coulomb collisions in fully ionized plasmas are found to mitigate the self-generated pressure anisotropies and promote the growth of the magnetic field.
\end{abstract}




\section{Introduction}

A population of ions drifting at super-Alfv\'enic speeds with respect to a magnetised ambient plasma can trigger various streaming instabilities, leading to the exponential growth of electromagnetic perturbations across a wide range of astrophysical environments \citep{kulsrudEffectWaveParticleInteractions1969,garyElectromagneticIonBeam1984,winskeDiffuseIonsProduced1984,bellTurbulentAmplificationMagnetic2004,amatoKineticApproachCosmicrayinduced2009,amatoOriginGalacticCosmic2014,cuiYoungSupernovaRemnant2016}. Depending on the plasma conditions, a mode not involving particle-wave resonances can grow \citep{winskeDiffuseIonsProduced1984} and drive turbulent magnetic field amplification that can greatly exceed the initial seed field \citep{bellTurbulentAmplificationMagnetic2004}. The significance of this non-resonant streaming instability (NR) was initially recognized for its role in the generation of diffuse ion distributions in the Earth's foreshock region, where reflected ions streaming back from Earth's bow shock encounter the incoming solar wind ions \citep{sentmanInstabilitiesLowFrequency1981, onsagerInteractionFinitelengthIon1991,akimotoNonlinearEvolutionElectromagnetic1993}. The discovery of its ability to strongly amplify the magnetic field has now made it a key ingredient in the diffusive shock acceleration of cosmic rays in supernovae remnants of particles up to PeV energies \citep{bellTurbulentAmplificationMagnetic2004,amatoKineticApproachCosmicrayinduced2009,bellCosmicRayAcceleration2013, marcowithMicrophysicsCollisionlessShock2016}. The NR instability has been studied numerically using a variety of computational techniques, including modified magneto-hydrodynamics (MHD) \citep{bellTurbulentAmplificationMagnetic2004,zirakashviliModelingBellNonresonant2008}, hybrid-Particle-In-Cell (PIC ions and massless fluid electrons) \citep{winskeDiffuseIonsProduced1984,akimotoNonlinearEvolutionElectromagnetic1993,haggertyDHybridRHybridParticleincell2019,marretGrowthThermallyModified2021}, full-PIC \citep{riquelmeNonlinearStudyBell2009, ohiraTwodimensionalParticleincellSimulations2009, crumleyKineticSimulationsMildly2019} and MHD-PIC \citep{baiMagnetohydrodynamicParticleinCellMethodCoupling2015,vanmarleMagneticFieldAmplification2018, mignoneParticleModulePLUTO2018} simulations.

The MHD-PIC method has received growing attention as it combines the kinetic treatment of the cosmic rays while retaining the advantage of modelling the background plasma as a magnetofluid, over large spatial and temporal scales. Neglecting kinetic effects in the background plasma however is not always justified. For example, in the hot plasmas of superbubbles or in the intergalactic medium, the background's ions thermal Larmor gyro-radius can become comparable to or larger than the unstable wavelengths. Under these conditions a kinetic treatment of the background ions population is necessary to account for resonance effects that can significantly reduce the growth of the NR instability \citep{revilleEnvironmentalLimitsNonresonant2008,zweibelEnvironmentsMagneticField2010,marretGrowthThermallyModified2021}. In addition, the development of significant ion pressure anisotropies in the background plasma have been observed in collisionless hybrid-PIC simulations \citep{marretGrowthThermallyModified2021}, and is found to be well described within the CGL adiabatic theory \citep{marretEnhancementNonresonantStreaming2022}. This suggests that the assumption of an isotropic scalar pressure, often employed in fluid models, may not be well-suited. Pressure anisotropies can be suppressed by particle collisions or by micro-instabilities such as the mirror and ion-cyclotron mode \citep{garyProtonTemperatureAnisotropy1976}, among other isotropization mechanisms. 
In this work we expand on previous studies \citep{marretEnhancementNonresonantStreaming2022} and investigate the interplay between magnetic field amplification and self-generated pressure anisotropies in plasmas with varying levels of collisionality.

The NR instability can be described by considering a modified MHD model \citep{bellTurbulentAmplificationMagnetic2004} with two components: a background plasma population, and a less dense cosmic ray population of protons, drifting with a velocity $u_{cr}$ parallel to an ambient magnetic field $B_0$. The perpendicular case is not considered here because of its slower growth rate \citep{matthewsAmplificationPerpendicularParallel2017}. The background plasma is electrically charged to satisfy the quasi-neutrality condition $n_m-n_e=-n_{cr}$, where the subscripts $m$, $e$ and $cr$ denote the main protons, electrons, and proton cosmic rays respectively. The background plasma carries a return current, produced by the electron drift relative to the main protons as $u_e=u_{cr}n_{cr}/n_m$, and which initially compensates the cosmic rays current. The electric field is obtained from an ideal Ohm's law with an additional term due to the presence of the low density drifting cosmic rays \citep{baiMagnetohydrodynamicParticleinCellMethodCoupling2015,vanmarleMagneticFieldAmplification2018}: $\vec{E}=-\vec{u}\times\vec{B}-\vec{j}_{cr}\times\vec B/en_m$, where $\vec j_{cr}=en_{cr}\vec u_{cr}$ is the current carried by the cosmic rays, initially opposed to the background return current, $e$ is the elementary charge, and $\vec u$ is the background fluid velocity. The essence of the instability can then be captured \citep{marretGrowthThermallyModified2021} by considering the simplified linearized momentum conservation for the background MHD fluid and Maxwell-Faraday's equations
\begin{align} 
\label{eq:momentum_heur}
\rho_0\dfrac{\partial\vec u_1}{\partial t} &= - \vec j_{cr}\times\vec B_1 \\
\label{eq:mag_heur}
\dfrac{\partial\vec B_1}{\partial t}&=(\vec B_0\cdot\vec\nabla)\vec u_1
\end{align}
where $\rho$ is the main plasma mass density. The numerical subscripts refer to initial and perturbed quantities. We are interested in electromagnetic modes and neglect density fluctuations. In addition, the cosmic rays current is taken to be constant and the magnetic tension is neglected. These approximations remain valid \citep{bellTurbulentAmplificationMagnetic2004} for unstable wave numbers in the range $k_{\min}<k<k_{\max}$, where 
\begin{align}
    k_{\min}&=\Omega_0/u_{cr}
    \\
    k_{\max}&=\frac{n_{cr}}{n_m}\frac{u_{cr}}{v_{A0}^2}\Omega_0
\end{align}
$\Omega_0=eB_0/m_p$ is the proton gyro-frequency, $m_p$ the proton mass, $v_{A0}=B_0/\sqrt{\mu_0\rho_0}$ the Alfv\'en velocity and $\mu_0$ the vacuum permeability. Larger scale perturbations with $k\leq k_{\min}$ are unstable to resonant modes, which require a fully kinetic treatment of the cosmic rays population \citep{amatoKineticApproachCosmicrayinduced2009}. The coupled equations \ref{eq:momentum_heur} and \ref{eq:mag_heur} show that the NR mode is driven by the cosmic ray current through the action of the magnetic force, $-\vec j_{cr}\times\vec B_1$, which produces fluid velocity fluctuations, $\vec{u}_1$, in the background plasma. The induced electric field, $-\vec u_1\times\vec B_0$, feeds back and enhances the initial magnetic field perturbation via Faraday's law, promoting its exponential growth. 
Considering a right-hand circularly polarized parallel propagating magnetic field perturbations of the form $B_1=\tilde{B}_1e^{i(kx-\omega t)}$, where the angular frequency $\omega=\omega_r+i\gamma$, $\omega_r\ll\gamma$ is taken to be positive, the growth rate for negative $k$ is given by 
Retaining the magnetic tension force in Eq. \ref{eq:momentum_heur} that dominates at the smallest scales yields a cutoff of the instability at $k=k_{\max}$, and a fastest growing mode with $\gamma_{nr}=k_{\max}v_{A0}/2$ for $k=k_{\max}/2$.
The magnetic field amplification can lead to pressure anisotropies due to adiabatic invariants conservation, which can in turn produce pressure gradients in the background plasma \citep{marretEnhancementNonresonantStreaming2022}.
The effect of pressure gradients on the instability cannot be captured by linear theory, since no transverse contribution remains in the linearized momentum conservation equation in the gyrotropic limit \citep{bellTurbulentAmplificationMagnetic2004}. We will show that non-linear pressure gradients effects can nonetheless modify the growth of the NR mode by opposing the magnetic force driving the magnetic field amplification.

\section{Numerical Setup}
We present the results of 1D and 2D simulations performed with the hybrid-PIC-MCC code HECKLE \citep{smetsr.HeckleHttpsGithub2011}, which solves the Vlasov-Maxwell system including Monte Carlo collisions, using a predictor-corrector scheme for the electromagnetic field, a non-relativistic Boris pusher \citep{borisAccelerationCalculationScalar1970}, and a first order interpolation scheme for current deposition on the grid. The main protons and cosmic rays protons are described with macroparticles following the PIC method, and the electrons are treated as a massless neutralizing fluid. This hybrid approach is well suited to study the kinetic, non-linear evolution of systems at the ions temporal and spatial scales while avoiding prohibitive computational time.
The density and magnetic field are normalized to their initial uniform values $n_0=n_m(t_0)$ and $B_0 = B(t_0)$. Times and lengths are normalized to the initial inverse proton cyclotron angular frequency $\Omega_0^{-1}$
and proton inertial length $l_0=v_{A0}\Omega_0^{-1}$.

We consider a population of cold main protons and electrons, traversed by a population of super-Alfv\'enic cosmic rays (protons) with a density $n_{cr}/n_m=0.01$. The initial drift velocity, $u_{cr}/v_{A0}= 50 $, is oriented parallel to the initial magnetic field $\vec B_0=B_0\vec e_x$. The cosmic rays current is not maintained externally, and thus decreases strongly upon the NR mode growth and saturation. The plasma and field quantities are initially homogeneous, and periodic boundary conditions were used in all directions. The simulation domain has a length $L_x=10^3 l_0$ where $l_0$ is the proton inertial length, discretized with $10^3$ cells for 1D simulations, and extended to $L_y = 2\times 10^2 l_0$ in the 2D case with the same resolution as in the $x$ direction. We used 200 macroparticles per cells in 1D simulations, 100 for each proton species. We used 600 macroparticles per cell for 2D simulations, 500 for the background protons and 100 for the cosmic rays. These values have been chosen to mitigate large numerical electric fields that may develop in the hybrid-PIC model in regions with few macroparticles per cell, while achieving numerical convergence and minimizing the numerical cost of the simulations. The electrons pressure is assumed to follow an isothermal behavior $P_e(t)=n_e(t)T_{e}(t_0)$ with $T_e(t_0)=T_m(t_0)=m_pv_{A0}^2$. Unless stated otherwise, the initial plasma-$\beta$ is $\beta_0=P_0/W_{B0}=2$ where $P_0$ is the initial, isotropic main protons pressure and $W_{B0}=B_0^2/2\mu_0$ is the initial magnetic field energy density. We also investigated the case $\beta_0=10$ by increasing the main protons initial temperature. The \textit{initial} main protons Larmor radius is then resolved with 2 cells for $\beta_0=10$ and 1 cell for $\beta_0=2$.
\begin{figure}
    \includegraphics[width=\columnwidth,trim=0 0cm 0 0,clip]{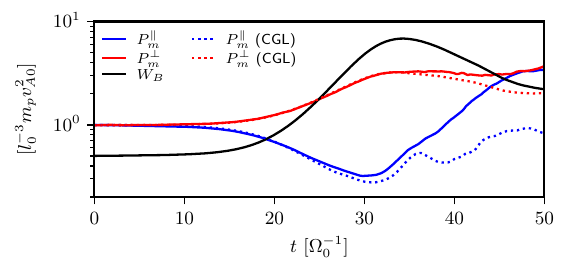}
    \caption{Evolution of the spatially averaged magnetic field energy density $W_B$ (solid black line) and of the spatial average of the $P_m^\perp$ and $P_m^{\smash[b]{\parallel}}$ components of the main protons pressure tensor calculated from the local macroparticles distribution (red and blue solid lines respectively), in a 1D collisionless simulation. The dotted lines show the CGL predicted pressures $P_m^\perp=P_0\rho B/\rho_0B_0$ and $P_m^{\smash[b]{\parallel}}=P_0 \rho^3B_0^2/\rho_0^3B^2$, calculated using the density and magnetic field extracted from the simulations.}
    \label{fig:figure_2}
\end{figure}

\section{Pressure anisotropies generation and interplay with micro-instabilities}
\label{sec:pressure_anisotropies}

To emphasize the relationship between the development of the NR mode and the emergence of pressure anisotropies, we display in Fig. \ref{fig:figure_2} the evolution of the magnetic field energy density ($W_B=B^2/2\mu_0$), together with the perpendicular and parallel components of the main protons' pressure tensor $P_m^\perp/P_m^{\smash[b]{\parallel}}$ with respect to the magnetic field, computed as the second order moment of the velocity distribution. Between $t=15$ and $t= 34 \,\Omega_0^{-1}$, the magnetic perturbations exhibit an exponential increase, accompanied by significant anisotropies, with a maximum spatially averaged value of $P_m^\perp/P_m^{\smash[b]{\parallel}} = 36.7$. Upon reaching saturation, the magnetic field energy density can be predicted by considering the energy exchange rates obtained within quasi-linear theory \citep{winskeDiffuseIonsProduced1984}, which indicates that the rate of energy gained by the magnetic field is half that of the cosmic rays' drift kinetic energy loss. Extrapolating this result to the non-linear evolution and supposing that the initial cosmic rays drift kinetic energy density is entirely depleted at saturation, one obtains $W_{B,\text{sat}} = W_{cr}/2$ with $W_{cr}=n_{cr}m_p u_{cr}^2/2$. This estimate is close to the simulations results $W_{B,\text{sat}} =0.55 W_{cr}$ obtained by averaging the magnetic field energy density over space, consistent with the fact that the cosmic rays current is not externally maintained and decreases strongly over the course of the simulation.

The development of pressure anisotropies can be described within the adiabatic CGL theory \citep{chewBoltzmannEquationOnefluid1956}, which may be interpreted as the conservation of the first and second adiabatic invariants in a slowly varying magnetic field \citep{kulsrudMHDDescriptionPlasma1983} and predicts a decrease in parallel pressure and an increase in perpendicular pressure in regions of amplified magnetic field for a constant density. Taking the electrons to be magnetized, considering an ideal Ohm's law (without the cosmic rays contribution), and neglecting heat fluxes and non-gyrotropic components of the pressure tensor, the main protons pressure parallel and perpendicular to the magnetic field may be expressed as
\begin{align}
\label{eq:cgl_para}
\dfrac{\d}{\d t}\left(\dfrac{P_m^{\smash[b]{\parallel}} B^2}{\rho^3}\right)=&\ 0 \\
\label{eq:cgl_perp}
\dfrac{\d}{\d t}\left(\dfrac{P_m^\perp}{\rho B}\right)=&\ 0
\end{align}
where $\d/\d t=\partial_t+\vec u_m\cdot\vec\nabla$ denotes the material derivative.
The advective part of the derivative may be neglected by integrating over the periodic simulation domain, and by neglecting the remaining velocity divergence term, which is a well verified assumption in the simulations. In this case the pressure components may be evaluated directly as $P_m^\perp/P_0=\rho B/\rho_0B_0$ and $P_m^{\smash[b]{\parallel}}/P_0=\rho^3B_0^2/\rho_0^3B^2$. 
These equations yield the well known effect that pressure anisotropies $P_m^\perp/P_m^{\smash[b]{\parallel}}>1$ can be produced in an amplified magnetic field at constant density, and are a good model for the simulations. In Fig. \ref{fig:figure_2}, we display the evolution of the pressure components $P_m^\perp$ and $P_m^{\smash[b]{\parallel}}$ predicted from the CGL model and calculated using the density and magnetic field from the simulation. The pressure anisotropies driven by the NR mode evolve according to the the CGL equations up to saturation ($t\approx 35\Omega_0^{-1}$).
This result confirms that the observed pressure anisotropies are not a byproduct of the restricted 1D and 2D geometry of the simulations. We note that the CGL equations do not account for the cosmic rays contribution in Ohm's law, but still describe well the anisotropies.

\begin{figure}
\centering    
\includegraphics[width=\columnwidth]{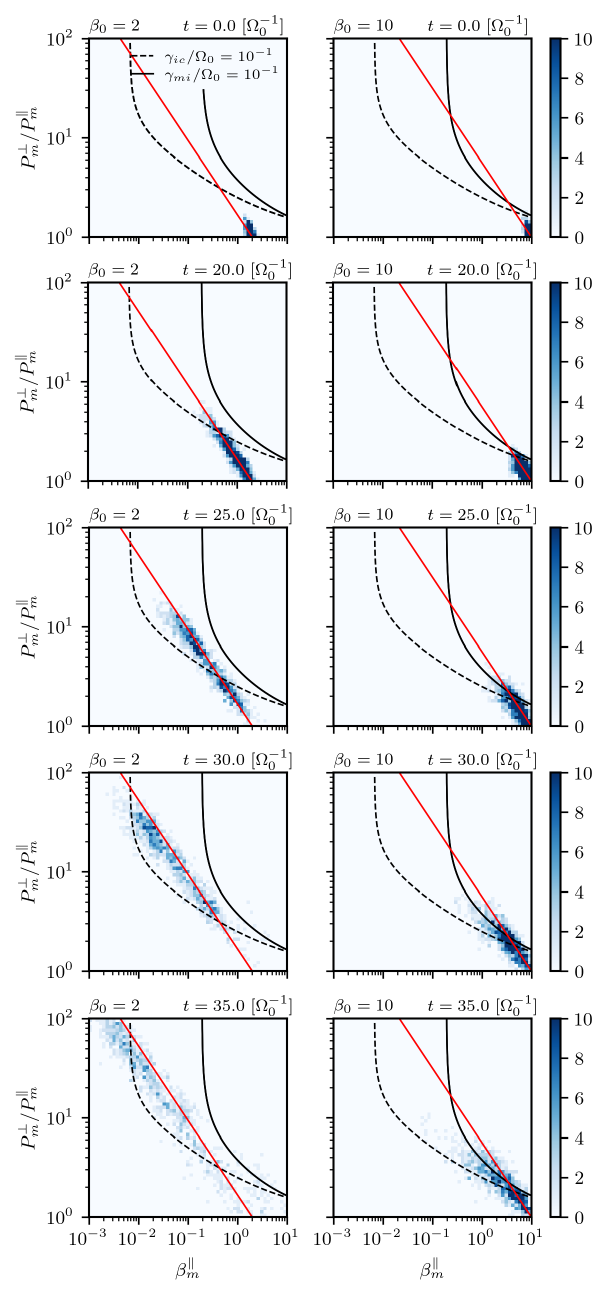}
\caption{Distribution (simulation cell count) of the ratio $P_m^\perp/P_m^{\smash[b]{\parallel}}$ as a function of $\beta_m^\parallel=P_m^{\smash[b]{\parallel}}/W_B$, obtained in 1D collisionless simulations with $\beta_0=2$ (left panels) and $\beta_0=10$ (right panels), at different times during the growth of the NR instability. The solid red line shows the anisotropy expected from incompressible CGL theory $P_m^\perp/P_m^{\smash[b]{\parallel}}=(\beta_m^\parallel/\beta_0)^{-3/4}$. The solid and dashed black lines indicate the thresholds for the mirror $\gamma_{mi}=10^{-1} \Omega_0$ and ion-cyclotron $\gamma_{ci}=10^{-1} \Omega_0$ modes respectively.}
\label{fig:figure_6}
\end{figure}

Pressure anisotropies $P_m^\perp/P_m^{\smash[b]{\parallel}}>1$ can play an important role in the plasma dynamics by triggering the growth of the ion-cyclotron ($ic$) and mirror ($mi$) micro-instabilities \citep{garyProtonTemperatureAnisotropy1976} upon reaching the threshold  $P_m^\perp/P_m^{\smash[b]{\parallel}}-1\geq S_p/(\beta_m^\parallel-\beta_p)^{\alpha_p}$ \citep{hellingerSolarWindProton2006}, where $S_p$, $\beta_p$ and $\alpha_p$ are fitting parameters determined numerically for a given $\gamma_\text{mi}$ and $\gamma_{ci}$. The micro-instabilities growth rate increases with larger $P_m^\perp/P_m^{\smash[b]{\parallel}}$, and is stabilized for smaller $\beta_m^\parallel$. While both instabilities feed on pressure anisotropies to develop, the mirror mode can be described at first order within a MHD framework, whereas the ion-cyclotron mode requires a kinetic description to correctly describe the wave-particle resonance leading to its growth \citep{southwoodMirrorInstabilityPhysical1993}.
To understand the role of micro-instabilities on the growth of the NR mode, we display in Fig. \ref{fig:figure_6} the distribution (cell count) of the ratio $P_m^\perp/P_m^{\smash[b]{\parallel}}$ as a function of $\beta_m^\parallel=P_m^{\smash[b]{\parallel}}/W_B$, for two collisionless simulations with $\beta_0=2$ and $\beta_0=10$. Fig. \ref{fig:figure_6} also shows the threshold anisotropy for the ion-cyclotron and mirror instabilities, obtained from linear kinetic theory assuming a homogeneous plasma with bi-Maxwellian populations \citep{hellingerSolarWindProton2006} and for a growth rate comparable to that of the collisionless NR mode ($\gamma_0=0.15 \Omega_0$).
Using the approximated forms of Eqs. \ref{eq:cgl_para} and \ref{eq:cgl_perp}, the anisotropy within the CGL theory can be described with a power law as
\begin{equation}
\dfrac{P_m^\perp}{P_m^{\smash[b]{\parallel}}}=\left(\dfrac{\beta_m^\parallel}{\beta_0}\right)^{-3/4} 
\label{eq:cgl_powerLaw}
\end{equation}
which is well recovered in the simulations with $\beta_0=2$ (Fig. \ref{fig:figure_6} left panels), indicating that the anisotropies are not constrained by micro-instabilities. In particular the mirror mode growth rate $\gamma_{mi}\sim 10^{-2} \Omega_0$ always remains below that of the NR mode measured in the simulation $\gamma_{nr}= 0.15 \Omega_0$. 

A different picture emerges for the simulation with $\beta_0=10$, obtained by increasing the main protons initial temperature, and shown in the right panels of Fig. \ref{fig:figure_6}. In this case the growth of the mirror mode competes with the NR mode ($\gamma_{mi}\sim\gamma$) for $P_m^\perp/P_m^{\smash[b]{\parallel}} \gtrsim 2$ and keeps the pressure anisotropies at the threshold values, below the expected power law. Similar results were obtained in 2D simulations. Although in both low and high-$\beta$ cases the ion-cyclotron instability has nominally a larger predicted growth rate than the mirror mode, and comparable or larger to that of the NR mode  for $P_m^\perp/P_m^{\smash[b]{\parallel}} \gtrsim 3$, it does not appear to be limiting the anisotropy; a similar behaviour has also been observed in the solar wind where pressure anisotropies above the expected ion-cyclotron threshold are frequently measured \citep{hellingerSolarWindProton2006}. In our case it may be due to the spatial inhomogeneity of the magnetic field amplification (see Sec. \ref{sec:collisions}), which renders the cyclotron resonance position dependent and impairs the growth of the ion-cyclotron mode \citep{southwoodMirrorInstabilityPhysical1993}. Studies of the growth of microinstabilities in the context of amplified magnetic field have also shown that the ion-cyclotron mode subdominance can be attributed to local departures from the Maxwellian distribution function \citep{riquelmeParticleincellSimulationsContinuously2015,waltersEffectsNonequilibriumVelocity2023} that is assumed when computing the ion-cyclotron mode linear properties, however we did not observe such departure in the simulations.

\section{Collisional non-resonant mode}
\label{sec:collisions}
In a collisional plasma, pressure anisotropies may be mitigated if collisions are sufficiently frequent to redistribute the energy in all directions of space. Expanding on previous works \citep{marretEnhancementNonresonantStreaming2022}, we investigate the two cases of first, a poorly ionized background where collisions of ions with a population of neutrals are dominant, relevant for example in the warm neutral medium \citep{recchiaGrammageCosmicRays2021}, and second, a fully ionized background plasma where ion Coulomb collisions are dominant, relevant to the interstellar medium. 
While intra-species Coulomb collisions conserve the momentum and energy \citep{trubnikovParticleInteractionsFully1965} of the population of particles of the same species, proton-neutral collisions do not, as protons loose their energy to the denser and colder background neutral population. This leads to important damping of the electromagnetic waves, including the NR mode \citep{fortezaDampingOscillationsIonneutral2007,revilleCosmicRayCurrentdriven2007}.

The proton-proton Coulomb collisions are implemented numerically using a Monte Carlo method which solves the Landau collisions operator by randomly pairing macroparticles in each cells, and calculating at each time step the associated scattering angle and post-collision velocities \citep{takizukaBinaryCollisionModel1977}.
The proton-neutral collisions are implemented following a hard-sphere model \citep{nanbuProbabilityTheoryElectronmolecule2000}. We considered proton-Hydrogen elastic collisions, and a small ionisation fraction such that the neutrals density and temperature are supposed to remain constant and uniform. At each time step a collision between a macroparticle and an hydrogen atom occurs for $r<n_n\sigma_{in}\Delta v\Delta t$, where $r\in [0,1]$ is a randomly generated number, $n_n$ is the neutral density, $\sigma_{in}$ the ion-neutral collision cross-section, $\Delta v$ the relative velocity and $\Delta t$ the time step. The cross-section collision energy dependency is obtained from \cite{krsticAtomicPlasmamaterialInteraction1999} assuming a neutral population temperature $T_n=1\ \text{eV}$ as typically observed in the warm atomic phase of the interstellar medium \citep{ferriereInterstellarEnvironmentOur2001}. 
The Coulomb and neutral collision frequencies, noted $\nu_0$, can be scaled with respect to the instability growth rate $\gamma$, allowing to probe the weakly ($\gamma_{nr}>\nu_0$) and strongly ($\gamma_{nr}<\nu_0$) collisional regimes of the NR mode in weakly or fully ionized plasmas. The electron resistivity is negligible for the range of Coulomb collision frequencies investigated, from $\nu_0=10^{-2}$ to $\nu_0=10^2\ \Omega_0$. Larger values were not investigated because of the prohibitively small numerical time steps required to resolve low energy collisions.

\begin{figure}
    \includegraphics[width=\columnwidth]{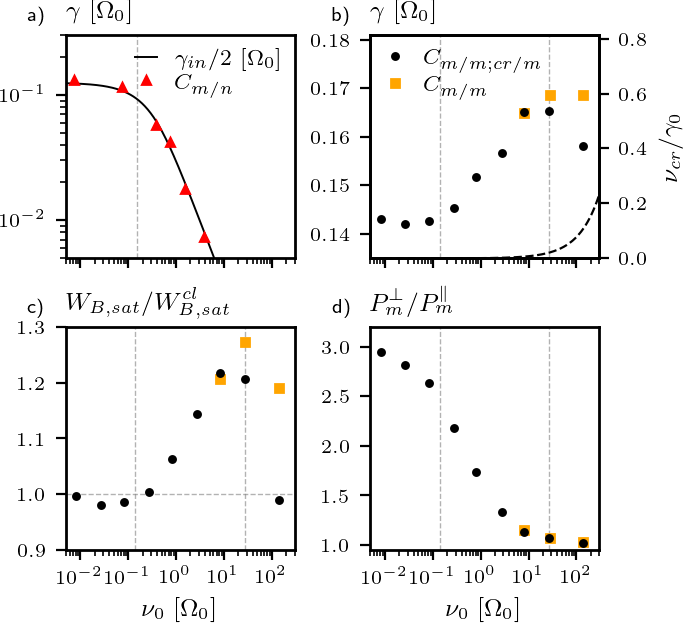}
    \caption{
    a) Measured magnetic field growth rate $\gamma$ ($C_{m/n}$, red triangles) as a function of the collision frequency, in 1D runs including proton-neutral collisions where $\nu_{0}=n_{n}\sigma_nv_{T0}$ is the main proton-neutral collision frequency, $n_n$ is the neutral density and $\sigma_{n}$ is the neutral collision cross-section. 
    The solid black line corresponds to the theoretical growth rate $\gamma_{in}/2$ considering $k=k_{\max}/4$. The collision frequency for which $\nu_{0}/\gamma_0=1$, where $\gamma_0=0.15 \Omega_0$ is the growth rate in the collisionless case, is indicated with the vertical dashed line and reported in the other figures.
    b) Magnetic field growth rate in 1D runs including Coulomb collisions between all protons populations ($C_{m/m;cr/m}$, black dots) and including Coulomb collisions between main protons only ($C_{m/m}$, orange squares), where $\nu_0=e^4n_m\ln{\Lambda}/4\pi m_p^2\epsilon_0^2 v_{T0}^3$ is the collision frequency among the main proton with $\ln{\Lambda}$ the Coulomb logarithm, and $v_{T0}=(T_m(t_0)/m_p)^{1/2}$ the thermal velocity. 
    The maximum growth rate for simulations with Coulomb collisions is indicated with the second vertical dashed line at $\nu_0=27\ \Omega_0$, and reported in the other figures. The dashed black line (bottom right) corresponds to the initial cosmic ray-main proton collision frequency $\nu_{cr}$, normalized to $\gamma_0$.
    c) Magnetic field energy density $W_{B,\text{sat}}=B_{\text{sat}}^2/2\mu_0$ at saturation, normalized to the value in collisionless simulations $W_{B,\text{sat}}^{cl}=6.8\ l_0^{-3}m_pv_{A0}^2$ and d) mean value of the ratio $P_m^\perp/P_m^{\smash[b]{\parallel}}$ averaged over the exponential phase of growth, for simulations including Coulomb collisions.}
    \label{fig:figure_3}
\end{figure}

Fig. \ref{fig:figure_3} presents the magnetic field growth rate $\gamma$ as a function of the reference collision frequency $\nu_0$, for simulations including a) neutral collisions and b) Coulomb collisions.
We find that the neutral collisions reduce the background fluid velocity perturbations, leading to a weaker induced electric field and consequently smaller growth rate and magnetic field amplification.
The damping of the instability by proton-neutral collisions can be studied by introducing a collisional drag term in the background fluid momentum conservation equation of the form $-\rho\nu_0(\vec u-\vec u_n)$ where $\vec u_n$ is the neutral fluid velocity. The growth rate in the case of a weakly ionized plasma is then given by $\gamma_{in}(k)=-\frac{\nu_{0}}{2}+\frac{1}{2}(\nu_{0}^2+4\Omega_0\frac{n_{cr}}{n_m}ku_{cr})^{1/2}
\label{eq:growth_neutral}$ \citep{revilleCosmicRayCurrentdriven2007}. This is plotted in Fig. \ref{fig:figure_3} a) considering $k=k_{\max}/4$ such that $\gamma_{in}(\nu_0=0)=\gamma_{nr}$, where $\gamma_{nr}=k_{\max}v_{A0}/2$. The growth rate dependency with collision frequency is well recovered in the simulations (red triangles). The magnetic field intensity has been integrated over the $k$ spectrum before measuring the growth rate, in order to reduce fluctuations due to the dynamic nature of the range of unstable wavenumbers inherent to the NR mode \citep{bellTurbulentAmplificationMagnetic2004,marretGrowthThermallyModified2021}. This gives an overall smaller growth rate than if only the fastest growing mode was observed and is seen with the offset by a factor 2 in the figure.

The simulations of a fully ionized collisional background (Fig. \ref{fig:figure_3} b) black dots) give an enhanced growth rate compared to the collisionless case $\gamma_0=0.15 \Omega_0$ for $\nu_0 > \gamma_0$. The increase is maximum for a collision frequency $\nu_0=27 \Omega_0$ two orders of magnitude larger than $\gamma_0$, yielding a growth rate $\gamma=0.17 \Omega_0$. This case will be referred as the collisional simulation in the following. The saturated magnetic field energy density $W_{B,\text{sat}}$ is displayed in Fig. \ref{fig:figure_3} c), and shows an increase up to $W_{B,\text{sat}}=8.7 l_0^{-3}m_pv_{A0}^2$ corresponding to $W_{B,\text{sat}}/W_{B,\text{sat}}^{cl}=1.3$ with $W_{B,\text{sat}}^{cl}$ the saturated magnetic field energy density in the collisionless case. The fastest growing wavenumber is not modified by the collision frequency, with unstable waves growing on scales of the order $\lambda\approx 4\pi k_{\max}^{-1}\approx 25 l_0$ in agreement with the linear kinetic theory prediction for a negligible background plasma temperature \citep{winskeDiffuseIonsProduced1984}. Because of the relatively large density of cosmic rays in the simulations, their collisions with the background protons become frequent for $\nu_0/\Omega_0\gg 10 $, leading to a rapid loss of their drift kinetic energy and to a lower magnetic field amplification.
This was verified by performing simulations where cosmic rays collisions were artificially suppressed (orange squares). In the case of completely collisionless cosmic rays, which is also more representative of the conditions found in space, the growth rate and magnetic field energy at saturation remain at the same level. 
Fig. \ref{fig:figure_3} d) shows the ratio $P_m^\perp/P_m^{\smash[b]{\parallel}}$ as a function of the collision frequency $\nu_0$. We find that the observed increase in the amplification of the magnetic field with $\nu_0$ is correlated to the gradual suppression of the pressure anisotropies for $\nu_0 \gtrsim 0.1 \Omega_0 \sim \gamma_0$.

\begin{figure}
\includegraphics[width=\columnwidth]{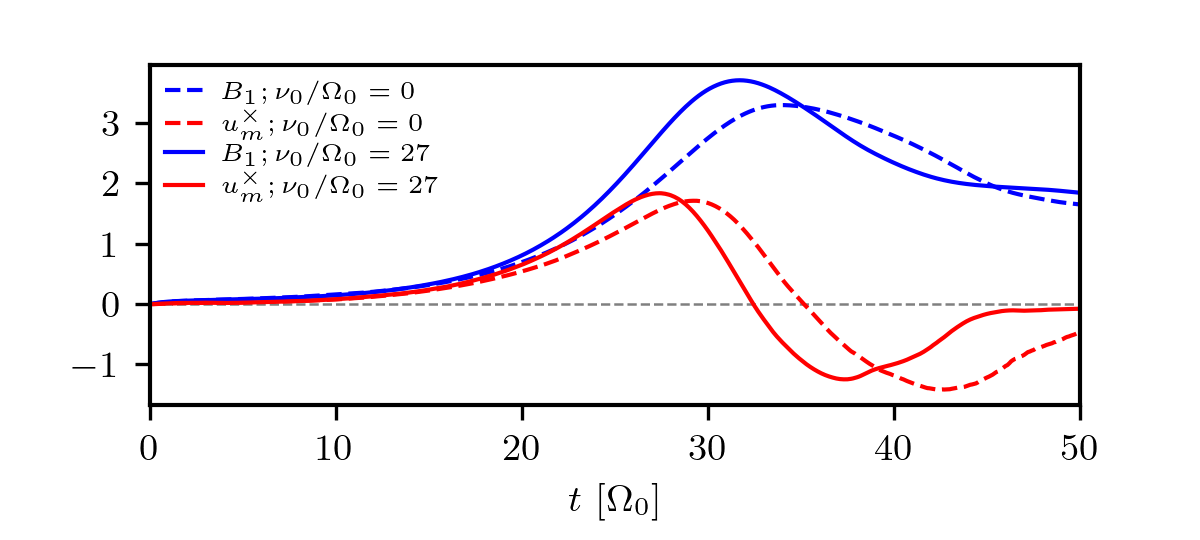}
\caption{
\textit{Perturbed magnetic field intensity $B_1$ in units of the initial magnetic field $B_0$ (blue) and background ion fluid velocity in the direction of the local $-\vec j_{cr}\times \vec B$ force $\mathbf u_m^\times$ in units of $v_{A0}$ (red), for 1D collisionless (dashed line) and collisional (solid line) simulations.
}}
\label{fig:figure_4}
\end{figure}

We interpret this result as follows. The pressure anisotropies generate spatial gradients of the pressure tensor along $B_0$, due to the helical spatial structure of the electromagnetic wave. Those pressure gradients affect the background plasma dynamics in the plane perpendicular to $B_0$ and oppose the magnetic force driving the NR mode. Suppressing these anisotropies promotes the growth of the magnetic field, however, we do not expect that a strong enhancement ($>100\%$) is possible since it would require the pressure gradients to overcome the cosmic rays magnetic force, which would prevent the growth of the instability altogether. The competition between the magnetic and the pressure gradient forces is illustrated in Fig. \ref{fig:figure_4} for 1D collisionless and collisional simulations. We compute the main protons fluid velocity component $\vec u_m^\times$ in the direction of the local $-\vec j_{cr}\times\vec B$ force, that is opposed by the pressure tensor gradients of the background protons $-\vec\nabla\cdot\vec P_m$. We find $|\vec j_{cr}\times\vec B|/|\vec\nabla\cdot\vec P_m|\sim 3$ during the exponential phase of growth in the collisionless case, which leads to a less efficient acceleration of the background fluid and to a proportionally smaller magnetic field amplification (Eq. \ref{eq:mag_heur}) compared to the collisional case. We verified that collisional viscous forces are negligible in the simulations. We note that in both cases the instability saturation occurs when the fluid velocity  $\vec u_m^\times$ changes sign, corresponding to the reversal of the longitudinal electric field $\vec E=-\vec u_m^\times\times\vec B$ that slows down the cosmic rays. This effectively halts the conversion of drift kinetic energy into magnetic field energy and saturates the NR mode \citep{marretGrowthThermallyModified2021}.

\begin{figure}
    \includegraphics[width=\columnwidth]{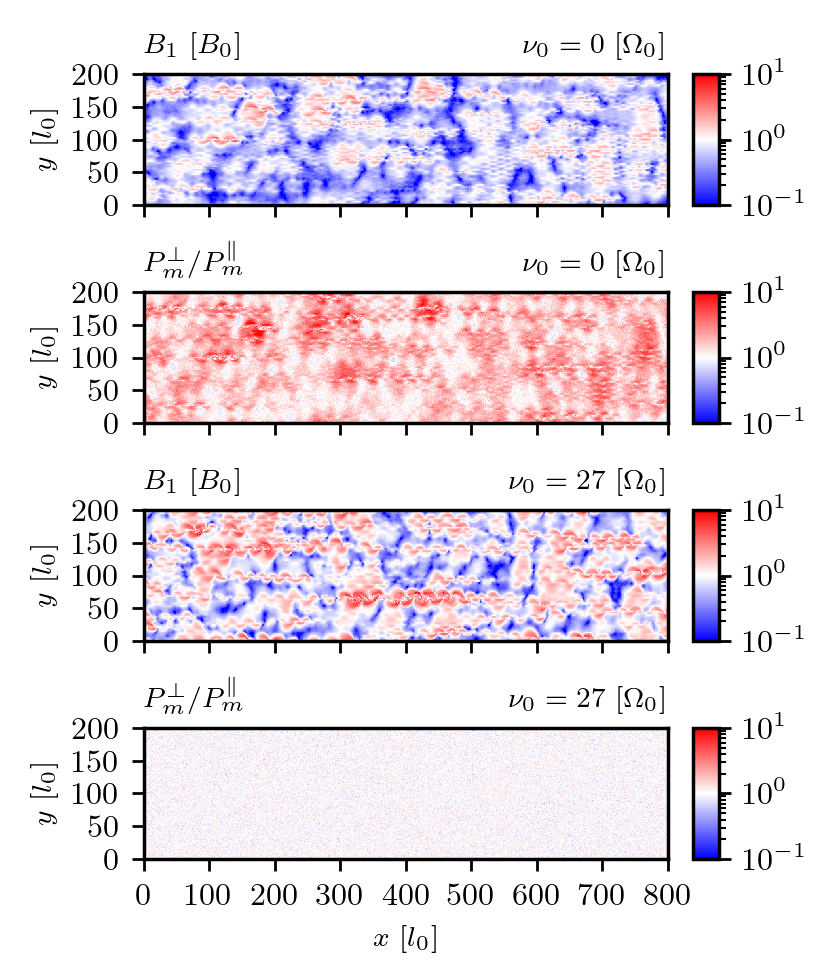}
    \caption{Maps of the perturbed magnetic field intensity $B_1$ and of the pressure anisotropy $P_m^\perp/P_m^\parallel$ spatial distribution, taken during the exponential phase of growth at $t=33\Omega_{0}^{-1}$ in 2D collisionless ($\nu_0= 0 \Omega_0$, two upper panels) and collisional ($\nu_0 =27 \Omega_0$, two lower panels) simulations.
    }
    \label{fig:figure_5}
\end{figure}

In 2D simulations, we find a growth rate of the magnetic field intensity averaged over space marginally larger ($\sim 1\%$) in the collisional case. The magnetic field is amplified similarly to the 1D runs, with an average magnetic field energy density ratio between the collisional and collisionless simulations of $W_{B,\text{sat}}/W_{B,\text{sat}}^{cl}=1.3$. This is illustrated in Fig. \ref{fig:figure_5}, which displays maps of the perturbed magnetic field intensity, together with maps of the pressure anisotropies $P_m^\perp/P^\parallel_m$, for 2D collisionless and collisional ($\nu_0=27 \Omega_0$) runs. We find that regions of magnetic field amplification are well correlated with the regions of large pressure anisotropies, further confirming our interpretation of the 1D simulations results.
In the collisional case, the anisotropies are suppressed ($P_m^\perp/P_m^\parallel\approx 1$) and the magnetic field is further amplified compared to the collisionless case. We note that the magnetic field amplification is spatially inhomogeneous, which may explain the absence of the ion-cyclotron mode in the simulations as mentioned in Sec. \ref{sec:pressure_anisotropies}.

\section{Discussion}
\label{sec:analytical_model}

We have presented 1D and 2D hybrid-PIC simulations of the non-resonant streaming instability including Coulomb collisions between protons and elastic collisions with neutrals. We have shown that the instability leads to an important anisotropic heating of the background plasma, well described by the adiabatic CGL theory and, depending on the initial plasma-$\beta$, constrained by the mirror micro-instability. The pressure gradients that are generated partially oppose the magnetic force and therefore the growth of the unstable waves. In fully ionized collisional plasmas, proton-proton Coulomb collisions reduce the pressure anisotropies, increasing the growth rate of the unstable waves and the amplification of the magnetic field. Simulations of poorly ionized plasmas, where proton-neutral collisions dominate, instead confirm the strong damping of the NR mode predicted by linear theory calculations \citep{revilleCosmicRayCurrentdriven2007}.

One may evaluate the Coulomb collision frequency required to mitigate the pressure anisotropies by comparing the NR mode anisotropic heating rate to the isotropization rate by collisions. The evolution of the anisotropy, $P_m^\perp-P_m^\parallel$, within incompressible CGL theory and including collisions can be expressed as
\begin{equation}
\dfrac{\partial }{\partial t}(P_m^\perp-P_m^\parallel) = \gamma (P_m^\perp+2P_m^\parallel) -\nu_0 \kappa P_0^{3/2} \dfrac{1}{\sqrt{\smash[b]{P_m^\parallel}}}
\label{eq:anisotropic_evolution}
\end{equation}
where the first term on the right hand side is the anisotropic heating rate due to the amplification of the magnetic field \citep{hunanaIntroductoryGuideFluid2019} by the NR mode, and the second term is the competing pressure isotropization rate due to Coulomb collisions \citep{trubnikovParticleInteractionsFully1965}. The magnetic field growth rate is defined as $\gamma\equiv(\d B/\d t)/B$, with $B=|\vec B|$; $\nu_0$ is the fundamental Coulomb collision frequency, and
\begin{equation}
    \kappa=3(2\pi^{1/2}A)^{-1}\left(-3+(A+3)\left[\tan^{-1}(A^{1/2})/A^{1/2}\right]\right)
\end{equation}
is a decreasing function of $A=P^{\perp}_m/P^{\parallel}_m-1$. 
Inserting the simplified Eqs. \ref{eq:cgl_para} and \ref{eq:cgl_perp} in Eq. \ref{eq:anisotropic_evolution}, 
one then obtains the level of collisionality necessary to stop the growth of the pressure anisotropies and reach steady state
\begin{equation}
\dfrac{\nu_0}{\gamma} = \dfrac{1}{\kappa}
\label{eq:nu_gamma_steady}
\end{equation}
Larger collisions frequencies, i.e. $\nu_0 / \gamma > \kappa^{-1}$, will start to strongly reduce the development of pressure anisotropies. The function $\kappa$ requires to calculate the parameter $A$, which may be inferred from the saturated magnetic field energy density prediction $W_B\approx W_{cr}/2$ obtained from quasi-linear theory \citep{winskeDiffuseIonsProduced1984}, such that $A=(W_{cr}/2W_{B0})^{3/2}-1$.
The above estimates is valid if the anisotropy is controlled only by the magnetic field amplification. However, pressure anisotropies with $P_m^\perp/P_m^{\smash[b]{\parallel}}>1$ can also drive the growth of the mirror mode. In the case where it competes with the NR mode, the maximum level of anisotropy will then be determined by the mirror mode threshold pressure anisotropy as presented in Sec. \ref{sec:pressure_anisotropies}. The rapid growth of the mirror mode also leads to important density fluctuations \citep{southwoodMirrorInstabilityPhysical1993} which would invalidate the incompressibility assumption in Eq. \ref{eq:anisotropic_evolution}, and thus make analytical estimates of $P_m^\perp$ and $P_m^\parallel$ near saturation unreliable, especially in environments with $\beta_0\gg 1$ where the mirror mode growth rate is maximum \citep{garyProtonTemperatureAnisotropy1997}. For this reason, we restrict our analysis to regimes where $\beta_0\lesssim 1$, such that the parallel plasma beta in the amplified magnetic field is small $\beta_m^\parallel\ll 1$, and the mirror mode remains subdominant.

For our simulation parameters, $\kappa^{-1}=7.3$, which agrees well with the range of collision frequencies, $\nu_0/\gamma>1$, for which pressure anisotropies are seen to be strongly reduced. Keeping the anisotropy small, say $P_m^\perp/P_m^{\smash[b]{\parallel}}-1< 0.1$, requires collision frequencies $\nu_0/\gamma>10^2$, which is again consistent with the values obtained in the simulations. The parameter $\kappa^{-1}$ can also be used to assess the importance of Coulomb collisions and of pressure gradients in various environments. Considering the situation of a supernova remnant (SNR) shock propagating at a velocity $u_{cr}=5\times 10^3 \,\text{km s}^{-1}$ in a fully ionized interstellar medium with $n_m=1 \,\text{cm}^{-3}$, $B=5 \, \mu\text{G}$, $T_m=10^{4} \, \text{K}$, and a cosmic rays flux $n_{cr}u_{cr}=5\times 10^{4} \, \text{cm}^{-2}\, \text{s}^{-1}$ \citep{zweibelEnvironmentsMagneticField2010}, we obtain $\beta_0=1.4$, $\gamma/\Omega_0=2.3\times 10^{-2}$, $A=33.0$ and $\kappa^{-1}= 6.8$, which is larger than the normalized proton Coulomb collision frequency \citep{trubnikovParticleInteractionsFully1965} $\nu_0/\gamma= 6.2\times 10^{-3}$. Under such conditions, pressure anisotropies will develop unimpeded and act to reduce the NR mode growth rate and saturated magnetic field. Proton-neutral collisions are not relevant in this case, however one may also consider the situation of a SNR propagating in a molecular cloud. One then finds that the NR mode is strongly damped by the collisions with the weakly ionized background neutrals, preventing magnetic field amplification by the instability \cite{revilleCosmicRayCurrentdriven2007}.

In the context of laboratory experiments, and considering parameters typical of high power laser-plasma interactions $u_{cr}=10^3 \text{km s}^{-1}$, $n_m=10^{19} \text{cm}^{-3}$, $B=0.2 \text{MG}$ and $T_m=10^6 \text{K}$, and a proton flux $n_{cr}u_{cr}=10^{26} \text{cm}^{-2}s^{-1}$, one finds $\beta_0=8.7\times 10^{-1}$, $\gamma/\Omega_0=3.6\times 10^{-1}$, $A=3.3$ and $\kappa^{-1}=5.6$, smaller than the collision frequency $\nu_0/\gamma=29.7$. In this case pressure anisotropies will be mitigated, enabling a faster growth of the magnetic perturbations.

Recent 1D numerical simulations of shocks including particle collisions have shown that the whistler waves excited upstream of the shock are strongly damped by Coulomb collisions \citep{nakanotaniCollisionalMagnetizedShock2022}. Although this may hinder the diffusive shock acceleration process by degrading the confinement of particles at the shock front, the opposite effect of enhancing the NR mode may point toward a different conclusion. Large scale hybrid-PIC simulations with a self-consistent kinetic description of the ion populations and including particle collisions will be necessary to assess the resulting cosmic rays acceleration efficiency.

\section*{Acknowledgements}

This work was performed using HPC resources from Grand Equipement National de Calcul Intensif- [Tr\`es Grand Centre de Calcul] (Grant 2019- [DARI A0060410819]), and was granted access to the HPC resources of MesoPSL financed by the Region Ile de France and the project EquipMeso (reference ANR-10-EQPX29-01) of the programme Investissements d'Avenir supervised by the Agence Nationale pour la Recherche. 

\section*{Data Availability}

The data underlying this article will be shared on reasonable request to the corresponding author.



\bibliographystyle{mnras}
\bibliography{references} 








\bsp	
\label{lastpage}
\end{document}